\documentclass[%
 reprint,
nofootinbib,
amsmath,amssymb,
aps,
prd,
showkeys,
]{revtex4-2}

\usepackage{graphicx, xcolor, orcidlink, comment, natbib}
\usepackage{dcolumn}
\usepackage{bm}
\usepackage{hyperref}

\DeclareMathOperator{\E}{\mathbb{E}}

\begin{document}

\title{Model-Independent Search Discards Faint Lensed-Pairs of Gravitational Wave Events in the Sub-Threshold Candidates of GWTC-4}

\author{Aniruddha Chakraborty\,\orcidlink{0009-0004-4937-4633}}
\email{aniruddha.chakraborty@tifr.res.in}
\author{Suvodip Mukherjee\, \orcidlink{0000-0002-3373-5236}}%
 \email{suvodip@tifr.res.in}
\affiliation{%
Tata Institute of Fundamental Research, 1, Homi Bhabha Road, Navy Nagar, Colaba, 
Mumbai 400005, India}%


\begin{abstract}
Gravitational lensing of gravitational waves (GWs) can produce multiple images in the geometric optics limit. These lensed GW images arrive at different times, are amplified by different magnification factors, and are shifted by constant phases. With current understanding, the occurrence of lensed events stands at a few per thousand events, and the number of GW detections is a few hundred with the ground-based detector network. However, with the inclusion of the sub-threshold events, the total number of detections crosses a few thousand. Therefore, a search that includes both types of events yields a higher chance of lensing detection. In this work, we carry out the first model-independent lensing search using a cross-correlation-based technique \texttt{GLANCE} over the entire volume of the GWTC-4 strain data, containing $\sim 90$ super-events $\sim 800$ sub-events forming a total of $\sim 11,000$ event pairs with a higher False Alarm Rate (FAR) event rate allowing to search deep in the noise dominated regime. We further conduct their spectrogram checks to inspect data quality, sky-map overlap of the interesting pairs, and a Bayesian parameter exploration of the sub-event to make a robust lensing detection. Although the search indicated four pairs of potential events with cross-correlation significance $\geq 2\sigma$, none were above $3\sigma$ at both the LIGO-Hanford and LIGO-Livingston detectors. This makes it possible to strongly rule out the presence of any statistically significant sub-threshold lensed GW event in GWTC-4. The null detection translates to an upper bound on the lensing detection rate to be $\leq$ 1.5/yr with inclusion of the sub-threshold event candidates. In the future, with more observation time, the detection of lensed GW can be possible from the current generation of GW detectors. 
\end{abstract}

\maketitle

\section{Introduction}

The emission of gravitational waves (GWs) is the way of energy dissipation from a massive system when the underlying spacetime undergoes violent changes \citep{Einstein:1915ca, PhysRev.117.306, 1975ApJ...195L..51H, Cutler:1994ys}. With a few hundred GW detections, the LIGO-Virgo-KAGRA (LVK) collaboration has probed compact object coalescences (CBCs) (with most of them from binary black hole (BBH) coalescences) in the universe up to a maximum redshift of $z \approx 1$ \citep{KAGRA:2013rdx, LIGOScientific:2014pky, PhysRevD.111.062002, LIGO:2024kkz, PhysRevD.102.062003, PhysRevLett.123.231107, PhysRevD.93.112004}. Owing to its low scattering cross-section with intervening matter \citep{Bishop:2024dej}, GWs offer insights about the high-energy, yet small-scale information about the compact objects in the early universe \citep{Maggiore:1999vm, santos2025gravitationalwavesearlyuniverse}. The GW detections from CBCs can be used to study several classes of studies of merging binary black holes (BBHs) properties \citep{KAGRA:2021duu, 2025arXiv250818083T}, tests of fundamental physics and general relativity \citep{LIGOScientific:2016lio, LIGOScientific:2018dkp, LIGOScientific:2019fpa, LIGOScientific:2020tif, 2026arXiv260319019T}, along with inferences of the cosmological parameters \citep{LIGOScientific:2021aug, 2025arXiv250904348T}- to mention a few.

Gravitational lensing is the bending of a light-like trajectory by a massive object \citep{1936Sci....84..506E, Bartelmann:2010fz}. Lensing amplifies the incoming flux of light, with differently amplified images arriving to us in different times \citep{1992grle.book.....S}. The mass distribution and the impact parameter of the source decides the specific effects imparted on the wave, in the geometric optics regime, given by $\rm \lambda_{gw} \ll R_{sch}$ \footnote{Here, $\lambda_{gw}$ is the wavelength of the GW and $\rm R_{sch} \propto M_{lens}$ is the Schwarzchild radius of the lensing object. Also, it can be written $f \gg t_d$ where $f$ is the frequency of the wave and $t_d$ is the time-delay between images as the lensing time-delay is proportional to the mass of the lensing object}, the images can be well-resolved spatially \citep{Takahashi:2003ix}. This regime, also known as geometric optics, provides a unique opportunity for the exploration of the high-redshift universe by enhancing any astrophysical signal, often not detectable by direct observations \citep{Smith:2017mqu, Xu:2021bfn, Vujeva:2025kko, Chen:2026qtu}.

Similarly to gravitational lensing of light, GWs can be deflected by massive objects along their trajectory \cite{Li:2018prc, 2023Univ....9..200G, Villarrubia-Rojo:2024xcj, Smith:2025axx}. In our previous works, with its 218 observed high-significance GW events up to the observation run in GWTC-4 \citep{LIGOScientific:2018mvr, LIGOScientific:2020ibl, LIGOScientific:2021usb, KAGRA:2021vkt}, we do not have evidence of any significant GW lensing detection \citep{Chakraborty:2025maj, Chakraborty:2025pxt}. Works performed by the LVK collaboration can be found here \citep{LIGOScientific:2021izm, LIGOScientific:2023bwz, lvk_gwtc4_lensing}. The current estimates hint at the possible lensing rate of GW events at $0.1-0.6\%$ \citep{Chen:2003uu, Robertson:2020mfh, Mukherjee:2021qam, Xu:2021bfn}. Therefore, with many more events in the future, the lensing of GW events is bound to be observed. Since the lensing probability (or what is called the optical depth of lensing) increases as the number of massive structures increases between the source and observer \citep{PhysRevD.101.123512, Mukherjee:2020tvr}, with the current generation of GW detector sensitivities (up to $z \approx 1$) the lensing event rate is limited to only a few lensed events per thousand unlensed ones.

The lensing optical depth increases with redshift; therefore, farther away GW sources have a higher probability of being lensed. However, the differential lensing optical depth also depends on the magnification of the lensing given by $dP/d\mu \propto 1/\mu^3$ \citep{PhysRevD.101.123512, Mukherjee:2020tvr} implies that sources with lower magnification are more likely to occur than the highly magnified ones. Whereas the highly magnified lensed GWs appear as super-threshold signals, the low-magnification GWs may fall in the sub-threshold lensed images \cite{Mukherjee:2021qam} \footnote{Super-threshold events are detections with a false alarm rate of $\leq$1/yr and sub-threshold events are detections with a false alarm rate of $>$1/yr.}. Therefore, looking in the entire catalog of GW events on the public domain \texttt{GraceDb}: Gravitational-Wave Candidate Event Database  \citep{2024EPJWC.29504022V}, it can be observed that there is an order of magnitude increase in the volume of the catalog when the sub-threshold events are included. Although many of these events can be noise mimicking as signals, there is a non-zero probability of obtaining GW events from these sub-threshold events. Comprising the sub-events, the catalog of events up to O4a observation contains about a thousand GW events. Due to aforementioned reasons, searching for a lensed pair of GW with the high-significant + low-significant GW events is of significant interest to the scientific community.

In this work, we \footnote{Throughout this work, "we" is used to solely refer to the authors of this paper.} carry out the first cross-correlation-based model-independent lensing search on pairs of high- and low-significant events to look for geometric-optics lensing pairs with \texttt{GLANCE}, first established in the work \citep{Chakraborty:2024net}. In this regime, a lensed GW pair is characterized by its magnifications, arrival times, and constant phase shifts. So, the key features of the unlensed GW, i.e., the phase, frequency, and amplitude evolutions, are unaffected when lensed. Two lensed GWs have nearly identical morphologies, and a cross-correlation-based strain-overlap strategy finds those lensed signals while suppressing the uncorrelated noise.

The work is divided into the following sections. In section \ref{sec:glance}, we describe the mathematical framework behind the working principle of \texttt{GLANCE} and how it employs a model-independent approach in detecting signals that have identical morphology. In section \ref{sec:method_and_cc}, we discuss the methodology for the selection of event-pairs and the results from performing cross-correlation on these pairs. The findings are also reported in the following section \ref{sec:results}. As a sanity check, we plot the spectrograms of the super-threshold and sub-threshold events and inspect the frequency evolution features in the signal. We also perform a Bayesian inference to characterize the source of the sub-threshold GW events from the significant pairs. In section \ref{sec:far}, we discuss the astrophysical contamination in lensing from GW sources in detecting the significant pairs. This would allow us to quantify the significance of any event pair as a lensed event, given the astrophysics of sources and their Poissonian nature in arrival. In the following section \ref{sec:rate}, we discuss the constraint on the upper bound on the lensing rate given the non-observation of any lensed events with current observations. We discuss a summarized form of this work and its conclusions and future implications in the final section \ref{sec:conclusion}.

\section{Brief Description of the cross-correlation Technique Deployed: \texttt{GLANCE}} \label{sec:glance}

\subsection{Mathematical structure of \texttt{GLANCE}}

The mathematical framework of \texttt{GLANCE} has been discussed in detail in our previous work \citep{Chakraborty:2024net}. For the sake of completeness, we briefly describe the key points here. \texttt{GLANCE} is a strain-level cross-correlation-based lensing search technique that does not rely on any astrophysical/waveform models for its operation. The key idea is to find the similarity between time-domain strain data segments and quantify the correlation strength as compared to noise.
To find the similarity between two data products, we calculate the cross-correlation defined by the following quantity,
\begin{equation}\label{eq:cc}
    a \otimes b = \frac{\langle a| b \rangle}{\sqrt{\langle a| a \rangle \langle b| b \rangle}} \hspace{0.25cm},
\end{equation}
where the inner product $\langle a| b \rangle$ for two time-series data is defined as:
\begin{equation}\label{eq:innerpdt}
    \langle a| b \rangle (t; t_d) = \frac{1}{\tau}\int _{t-\tau/2} ^{t+\tau/2}  a(t') b(t' + t_d)dt' \hspace{0.25cm}.
\end{equation}
The cross-correlation being a normalized quantity, it ranges between $-1$ and $+1$. In reality, the data are not a continuous variable of time; they are sampled at specific time-instances, and the above Eq. \ref{eq:innerpdt} can be obtained as

\begin{equation}\label{eq:discreetinnerpdt}
    \langle a| b \rangle (t_i; t_d) = \frac{1}{N}\sum_{t' =t_i-t_{i-N/2}}^{t' =t_i + t_{i+N/2}}a(t')b(t'+t_d),
\end{equation}
where, we use $dt' = \tau/N $. In general, we can describe any two time-series data $d(t)$ to consist of two components: signal $s(t)$ and noise $n(t)$ as, 
\begin{equation}
    d(t) = s(t) + n(t) \hspace{0.25cm}.
\end{equation}
The cross-correlation between two data-segments $d_1(t)$ and $d_2(t)$ is given by,
\begin{equation}
    d_1 \otimes d_2 = s_1 \otimes s_2 + n_1 \otimes n_2 + s_1 \otimes n_2 + n_1 \otimes s_2
\end{equation}
Given that the cross-correlation is the norm-weighted mean of the time-series segments, we can use the relations of expectation values as follows

\begin{eqnarray}
    a \otimes b \propto& \langle a | b \rangle \hspace{0.25cm}\text{(up to normalization)}, \nonumber \\
    =& \E [a \cdot b], \nonumber \\
    =& \rm{Cov}(a, b) + \E[a]\E[b].
\end{eqnarray}

For uncorrelated signals, we can show that $\E [a \cdot b] =\E[a]\E[b]$ \footnote{$\E [a \cdot b] = \int \int abp(a, b) dadb= \int ap(a)da \int b p(b) db = \E[a]\E[b]$ under the approximation that $p(a, b)$ can be decomposed into $p(a)p(b)$ for independent variables.}. For noise segments with a mean of zero, we have $\E[n_1] = 0$ and $\E[n_2] = 0$. Therefore, when the noise segments are independent, $n_1 \otimes n_2 \propto \E[n_1 \cdot n_2] = \E[n_1] \E[n_2] \rightarrow 0$ is the same as $\tau \rightarrow \infty$ (from the law of large numbers). From similar arguments $s_1 \otimes n_2 \propto \E[s_1 \cdot n_2] = \E[s_1] \E[n_2] \rightarrow 0$ and $s_2 \otimes n_1 \rightarrow 0$. Two signals $s_1$ and $s_2$, if they have a similar phase and frequency evolution, are correlated in nature. This results in $\rm{Cov} (s_1, s_2) \neq 0 $, although the signals may have $\E[s_1] = 0, \E[s_2] = 0$. Therefore, the cross-correlation between two similar signals is non-zero. However, if the signals are independent and uncorrelated, we have $\rm{Cov} (s_1, s_2) = 0$. Therefore, in a concise way, we can write that

\begin{equation}
d_1 \otimes d_2 =
\begin{cases}
\neq 0, & \text{if signals have similar phase evolution}, \\
= 0, & \text{else}.
\end{cases}
\end{equation}

In the case where there is no noise, for two identical signals, we have $ d_1 \otimes d_2 = 1$. With the presence of uncorrelated noise, it will not show a true overlap, thus becoming a value less than unity. For the same argument, two non-correlated signals should cross-correlate to zero under no noise. However, the presence of noise makes it deviate from zero. Thus, to find out which signals are correlated from a statistical point of view, we need to quantify how well we can detect them in terms of a cross-correlation signal-to-noise ratio (SNR). We define the cross-correlation SNR as,
\begin{equation}
    \rho_{cc}(t; t_d) = \frac{D_{12}(t; t_d)}{\sigma_{12}} \hspace{0.25cm},
\end{equation}
where, the cross-correlation on the data $D_{12}(t) = d_1(t) \otimes d_2 (t+ t_d)$ and $\sigma_{12}$ is the standard deviation of the cross-correlations performed between the segments where no signal is found, i.e., the noise segments $n_1(t') \otimes n_2(t' + t_d)$.

When a GW is lensed in the geometric-optics way, the lensed GWs are enhanced, phase-shifted, and time-delayed. The amplification factor (in the frequency domain) is given by

\begin{equation}\label{eq:sl_amp}
F(f)=\sum_j\left|\mu_j\right|^{1 / 2} \exp \left[2 \pi i f t_{d, j}-i \pi n_j\right] \hspace{0.2cm},
\end{equation}
where $j$ denotes the $j$-th image, $\mu_j$ is the magnification of the image, $t_{d, j}$ is the time-delay and $n_j$ is called the Morse phase factor, which can take values $\{0, 1/2, 1\}$ depending on specific conditions.

Therefore, the lensed GWs in the time domain can be written as

\begin{equation}\label{eq:sl_gw}
\phi^{ \rm lensed} (t, \vec{r})=\sum_j\left|\mu_j\right|^{1 / 2} \phi^ {\rm unlensed}\left(t-t_{d, j}, \vec{r}\right) \exp \left[-i \pi n_j\right] \hspace{0.2cm},
\end{equation}

where $\phi^{\rm unlensed}$ and $\phi^{\rm lensed}$ are the unlensed GW amplitude and lensed GW amplitude at a position $\vec{r}$.

Therefore, two gravitationally lensed counterparts have similar the morphologies of the signals \footnote{This is strictly valid in the geometric-optics regime, where the signals $s_1(t)$ and $s_2(t)$ are same up to a delay time difference $t_d$, a Morse phase $\Delta \phi$ and a magnification $\sqrt{\mu_{1}/\mu_2}$}, cross correlation between the data segments is expected to pick up the commonness between them as long as the noise in those segments is uncorrelated. In this case, we find the cross-correlation SNR at the time of coalescence $t_c$ with variable time-delays $t_d$'s based on which segments are intended to be checked for the similarities. For a more detailed explanation of the methodology applied, see here \citep{Chakraborty:2024net}.

\subsection{Salient Features of \texttt{GLANCE}}

In this section, we focus on the features of \texttt{GLANCE} and what makes it stand out from the existing methods for detecting lensing.

\begin{enumerate}
    \item \textbf{Usage of raw strain information:} Subjecting any data to the models can introduce biases and systematics depending on the approximations made. In \texttt{GLANCE}, we do not work with posteriors or any other derived product; instead we directly work with strains \footnote{We use calibrated, cleaned strain strain data from \texttt{GWOSC}.} from the detectors \citep{LIGOScientific:2019lzm, 2023ApJS..267...29A, 2025arXiv250818079T}, which are raw data products from the GW observatories. This makes \texttt{GLANCE} very immune to uncertainties from intermediate steps.
    \item \textbf{Application of phase information:} The evolution of the GW phase is a key piece of information to understand the masses and spins of the BBH system. In \texttt{GLANCE}, we used cross-correlation at the aligned phase level. Therefore, any mismatch in the two evolving GW phases makes the cross-correlation fall to zero and two lensed signals with identical phase evolution are picked up.
    \item \textbf{No dependence on models:} Current existing lensing detection methods rely on either the assumption of the GW waveform model or the gravitational lens model. In \texttt{GLANCE}, we  rely neither on the assumption of mass distribution of the lens nor on the modeling of the GW phase and amplitude. Therefore, the claim of lensing detection is very robust using this method.
\end{enumerate}

\subsection{Comparison with existing frameworks}

In the fourth version of the Gravitational Wave Transient Catalog (GWTC), the LVK collaboration implemented a few techniques to detect lensing in the multi-image and single distorted image regime \citep{2025arXiv251216347T}. In the multiple lensing regime, the collaboration has a few super-threshold lensing and a few sub-threshold lensing detection tools.

\begin{enumerate}
    \item \textbf{Super-threshold multi-image:} (a) \textbf{Posterior overlap} \citep{Barsode:2024zwv}: Relies on the detection of lensed pairs by calculating the overlap between the inferred source parameters of the two GW events. (b) \textbf{PHAZAP} \citep{2024ascl.soft06022E}: Checks the consistency of the evolution of the reconstructed GW phase in the mode (l, m) = (2,2). (c) \textbf{Golum} \citep{Janquart:2023osz}: Takes the posterior of image-I as input and links them to image-II and samples the lensing parameters, i.e., time-delay, magnification, and Morse phase factor, to find whether these 3 parameters converge. (d) \textbf{Hanabi} \citep{PhysRevD.107.123015}: Performs a joint PE for the two events, taking a single set of BBH parameters and the aforesaid 3 lensing parameters with SIS lens.
    \item \textbf{Sub-threshold multi-image:} (a) \textbf{PYCBC search} \citep{PhysRevD.102.084031}: Generates the best-fit template to a super-threshold GW signal and tries to match it among the sub-threshold images. (b) \textbf{TESLA-X} \citep{Li:2023zdl}: Finds overlapping sky-map GW events and constructs a template bank from the super-threshold image to search for the sub-threshold image signal.
    \item \textbf{Single distorted images:} (a) \textbf{Type-II lensing} \citep{Janquart:2023osz}: Images with Morse-phase of $\pi/2$ are detected in the framework of GOLUM, mentioned before. (b) \textbf{Search for wave-optics detections} \citep{Wright_2022}: Uses a point-mass lens model to jointly characterize the source and lens parameters in a Bayesian framework. (c) \textbf{Phenomenological searches} \citep{PhysRevLett.134.151401}: Searches for lensed GWs with very high and near equal magnifications happening near caustics curves of the source plane.
\end{enumerate}

In contrast to the other existing methods, we emphasize that \texttt{GLANCE} \citep{Chakraborty:2024net, Chakraborty:2024mbr} is the first detection method that does not rely on any assumption of the source, the waveform model, or the lens. In our previous studies \citep{Chakraborty:2025maj}, show that the conditions for a lensing detection in our analysis are so stringent that so far only one event has appeared as a wave-optics lensing candidate \footnote{Further tests revealed the event to be not significant as a wave-optics lensed event, shown in the same work.}. Thus, the potential of \texttt{GLANCE} to claim lensing in a robust way is now established, and these strain level cross-correlation studies with different approaches are followed up see \citep{Seo:2025dto, kopty2026}. We find that the findings from the LVK analysis and our analysis are consistent: the lensing of \href{https://gwosc.org/eventapi/html/GWTC-4.0/GW231123_135430/v2/}{GW231123\_135430} is supported least by \textbf{NRSur7dq4} \citep{Varma:2019csw} and slightly more in \texttt{IMRPhenomXPHM-SpinTaylor} \citep{colleoni2024}. See the works here \citep{lvk_gwtc4_lensing, Chakraborty:2025pxt}.

\section{Methodology and results from cross-correlating super-threshold events with sub-threshold} \label{sec:method_and_cc}

\begin{figure*}
    \centering
    \includegraphics[width=0.8\linewidth]{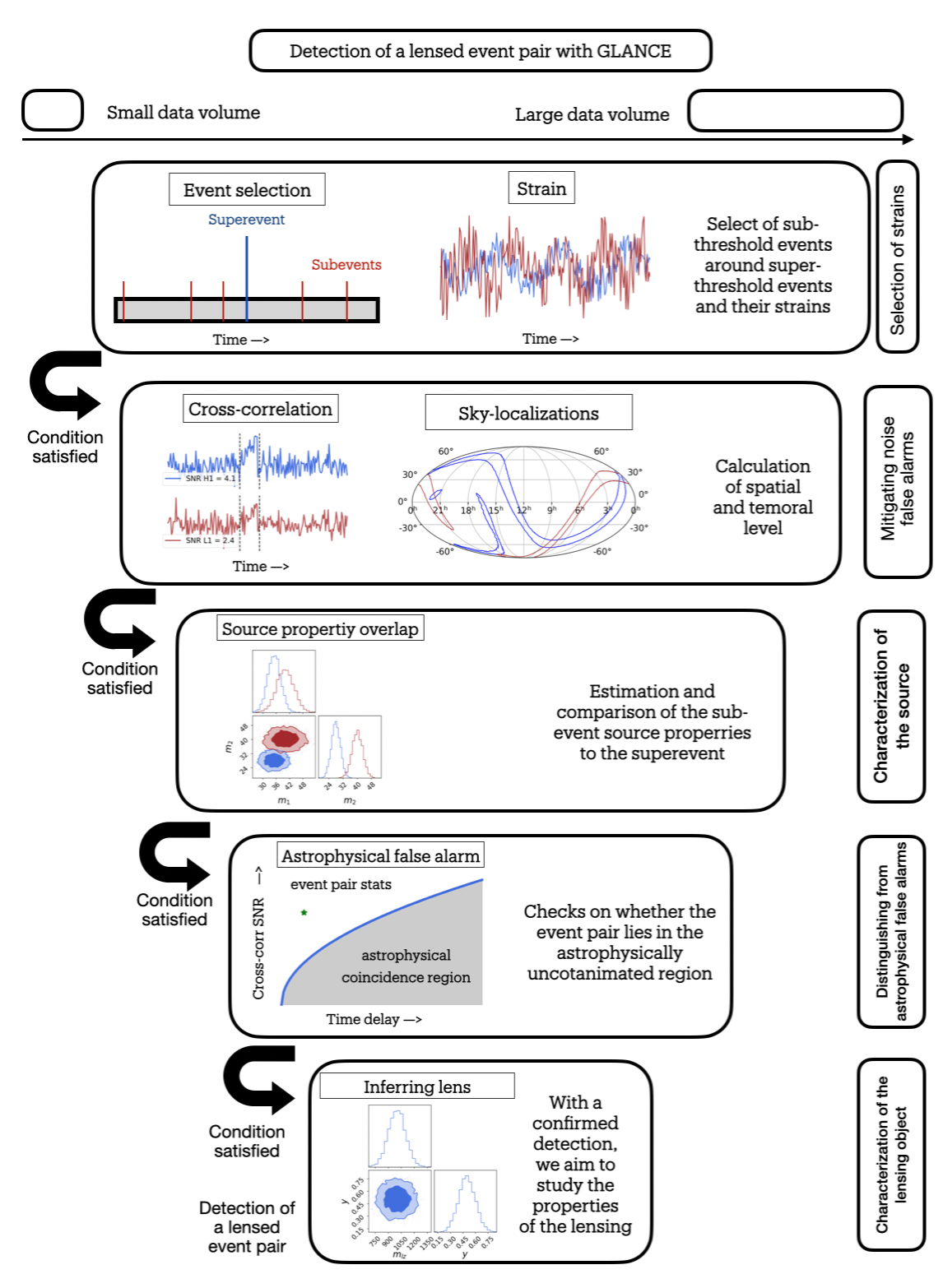}
    \caption{In this figure we present a schematic diagram for the lensing detection method deployed in this analysis. We first select the sub-threshold events around a super-threshold event with a cut on the astrophysical false alarm rate on those sub-events. We align each pair of events in time and perform the cross-correlation between the strains. If the cross-correlation SNR is above two in all detectors, we select those pairs. We find the overlap between the sky-localizations of the interesting pairs to determine whether they arrive from the same part of the sky. Finally, we check the posteriors of the super and sub-events if they contain overlaps in their sky-maps. If and only if the pair of events passes all these conditions, we qualify them as possible lensed events.}
    \label{fig:outline}
\end{figure*}

In GW observatories \citep{PhysRevLett.116.131103, Harry:2010zz, VIRGO:2014yos, PhysRevLett.123.231108, Virgo:2022ysc, Luck:2010rt, 2014CQGra..31v4002A, Dooley:2015fpa, KAGRA:2020tym, PhysRevD.88.043007, Somiya:2011np}, we have strain data as a function of discrete timestamps, sampled at 4096 Hz \citep{LIGOScientific:2019lzm, 2023ApJS..267...29A, 2025arXiv250818079T}. This allows any feature up to a Nyquist frequency of 2048 Hz to be observed. However, due to the photon shot noise dominance in the detector at high frequencies, current LVK detectors are only sensitive to $f \approx$ 1000 Hz \citep{LIGOScientific:2014pky}. On the other side of the spectrum, in the lower frequencies below $\approx$ 20 Hz, the data are contaminated by seismic noise. Therefore, we apply a whitening and band-passing (within $20$ Hz $-512$ Hz) on the data \citep{Finn:1992wt, Finn:1992xs, Cutler:1994ys, LIGOScientific:2019hgc, Flanagan:1997sx, Allen:2004gu, Allen:2005fk} to find GW signals in the frequency range where the noise strength (estimated in terms of the noise amplitude spectral density \footnote{The noise power spectral density is defined as: $S_n(f) = \frac{|\tilde{n}(f)|^2}{T}$ where T $\rightarrow \infty$. Amplitude spectral density, $A_n(f) = \sqrt{S_n(f)}$}) is lower than that of the signal.

\begin{figure*}
    \centering
    \includegraphics[width=\linewidth]{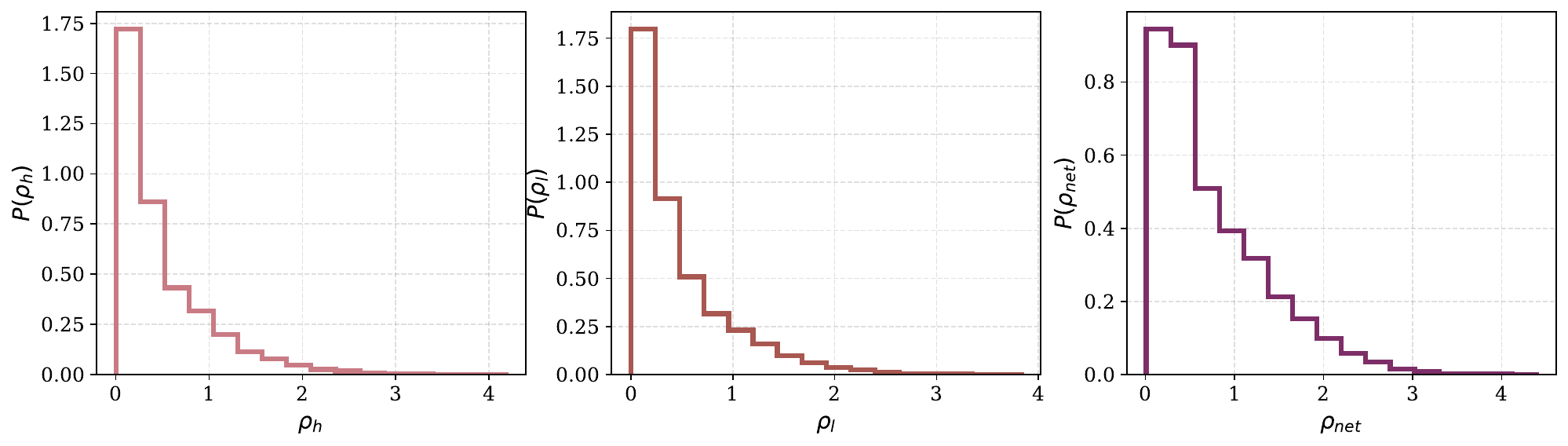}
    \caption{In this figure, we show the distribution of the cross-correlation SNRs at the H1 detector (left panel) and at the L1 detector (middle panel). The events that have at least an SNR of 2 at both detectors are subjected to further scrutiny. The network cross-correlation SNR distribution for these events is shown in the right panel.}
    \label{fig:snr_dist}
\end{figure*}

\subsection{Methodology applied for cross-correlation search}

We list down the steps for the methodology deployed for the selection and deployment of cross-correlation between the super-threshold and sub-threshold events. The approach is similar to our previous work \citep{Chakraborty:2025maj} and is presented through a schematic diagram in figure \ref{fig:outline}

\begin{enumerate}
    \item \textbf{Selection of high significance events:} We select high-significant events with an FAR $\leq$ 1/yr from the GWTC-4 catalog, appearing at a GPS time $t_{\rm event, 1}$. We also select only the events that are observed by at least a two-detector network.
    \item \textbf{Selection of low significance events:} We pick all low-significance events, that appear at the GPS time $t_{\rm event, 2}$ such that they lie within the time-window [$t_{\rm event, 1} - t_{\rm lensing}$, $t_{\rm event, 1} + t_{\rm lensing}$]. This event is chosen to have an FAR $\geq 1/\rm yr$ and is observed by at least two detectors. Since high time-delays are very prone to lensing false alarms from astrophysical sources, we have set $t_{\rm lensing}$ = 1 month. To remove the non-terrestrial origin events, we apply a false alarm rate cut-off for the sub-events need to be lower than 1/day. The $t_{\rm lensing}$ is motivated from our previous work on false alarm contamination in lensing detection \citep{chakraborty2026falsealarmratesdetecting}.
    \item \textbf{Time alignment of the signals:} We select the best-fit R.A. - Dec position of the high-significant event on the sky. For this sky position, we estimate the delay time between detectors in two different instances $t_{\rm event,1}$ (for a high-significance event) and $t_{\rm event, 2}$ (for a low-significance event). The delays between the detectors are used to align the signals. 
    \item \textbf{Cross-correlation of event strains:} We perform a strain level cross-correlation between the highly significant event strain and the low-significant event strain in both detectors. Then we apply a signal duration-wide averaging of the signal. This is compared with pure noise cross-correlation to obtain the cross-correlation signal-to-noise ratio $\rho_{\rm lensing}$.
\end{enumerate}

\subsection{Detection Conditions for a lensing pair with \texttt{GLANCE}} 

\begin{figure*}
    \centering
    \includegraphics[width=0.8\linewidth]{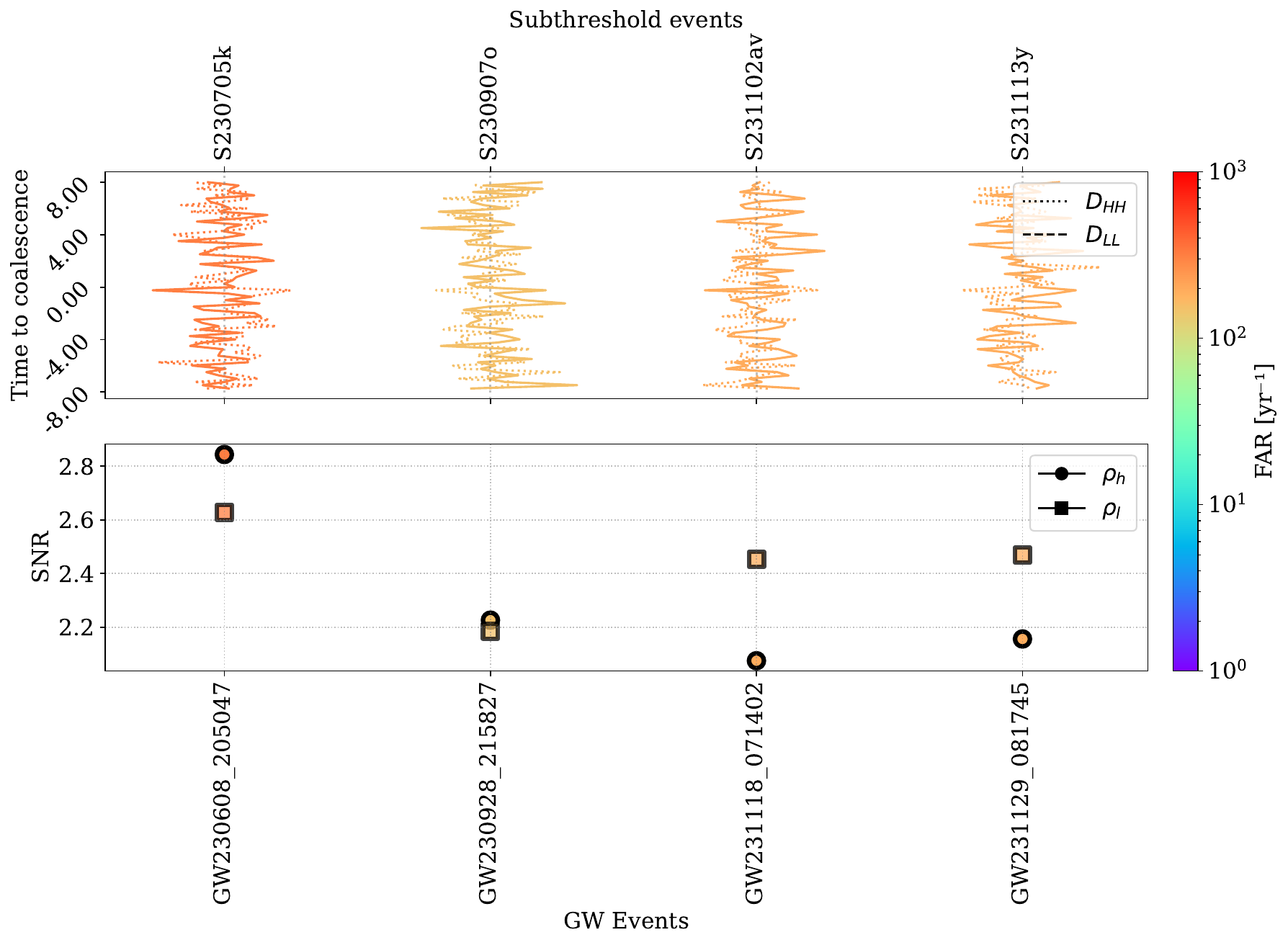}
    \caption{In this figure, we show the cross-correlation SNR for a cross-correlation timescale of 1/8s. 4 pair of events show up with a cross-correlation SNR threshold of 2 at both H1 and L1 detectors. However, the super-threshold signals all have an in-band duration between 0.1-0.2s. These short-duration signals cannot be characterized for lensing as the cumulative cross-correlation does not happen for a significant duration of time.}
    \label{fig:snr_0.125s_16s}
\end{figure*}

We find a total of 85 high-significance GW events op to O4a observed by a two-detector network \citep{KAGRA:2021vkt, 2025arXiv250818082T}, combined with a total of 780 low-significance two-detector events, producing a total number of 11,207 pairs when combined with the low-significance events. The cross-correlation SNR of these events is presented in \ref{fig:snr_dist}. The first two panels show the cross-correlation SNR distribution obtained for these event pairs from the H1 ($\rho_h$) and L1 ($\rho_l$) detectors, respectively. In the third panel, we show the network SNR $\rho_{\rm net} = \sqrt{\rho_h^2 + \rho_l^2}$. We perform the cross-correlation between these pairs, with the cross-correlation timescales of 1/8s \footnote{The results with other cross-correlation timescales of 1/16s are discussed in the appendix \ref{app:cc_timescale}}. To select an event pair as a lensing candidate, we set the following conditions:

\begin{enumerate}
    \item \textbf{Persistent cross-correlation signal:} The event pair cross-correlation should show a persistent cross-correlation over the duration of the signal. Since the detections are based on the merger time \footnote{This merger time is defined as the instance of the peak amplitude in the signal.}, any two signals aligned by their merger time always show up in cross-correlation due to the overlap in the final few cycles, but they deviate as the phase evolution is traced backwards. Thus, we accumulate the net cross-correlation signal over this duration and compare the signal strength above noise. We pick all possible event pairs, showing a cross-correlation SNR above 2. These are the interesting candidates that we will follow up on further.
    \item \textbf{Data quality inspection through spectrograms:} From the point of view of checking the data quality near the event, we plot the spectrograms of the super-threshold and sub-threshold events to visually inspect the frequency evolution track of the super-threshold and sub-threshold GW events. This helps us to rule out potential correlated noise artifacts in the data by manual inspection.
    \item \textbf{Overlapping sky-localizations:} For the interesting event pairs, we estimate the overlap in their sky-localizations. We select all those event pairs that show an overlap of their sky-localization at the 95\% credible interval.
    \item \textbf{Rejection of short signals:} We reject signal pairs if the duration of the super-threshold signal is shorter than $0.5$ s. This allows us to only study and select those pairs which have accumulated a cross-correlation signal over sufficient time and not just near the merger instance where all signals have significant overlap \footnote{This can be translated to matched-filter searches performed over a frequency range $\rm f_{min}$ to $\rm f_{max}$. If the frequency range spent by the signal is small, as in GW231123, GWs of very different source properties can match the signal \citep{LIGOScientific:2025rsn}. However, if the frequency range of the signal is sufficiently broad and the frequency evolution is captured by the detectors, the best-fit is unambiguous and the source parameters are constrained up to a small part of the parameter space.}. This has been discussed in detail in our previous work \citep{Chakraborty:2025maj}.
    \item \textbf{Inference of the low-significant source:} For these event pairs passing the tests before, we estimate the source parameters of the sub-threshold GW signal. We then compare the mass and spin inferences of the super-sub pair. Although this introduces a model dependence, it would allow for a one-to-one check, to confirm the candidature as a lensed pair. 
\end{enumerate}

\section{Results on lensed-pair searches}\label{sec:results}

In this section, we discuss the results in light of the checklist points we mentioned in the previous section.

\begin{figure*}
    \centering
    \includegraphics[width=\linewidth]{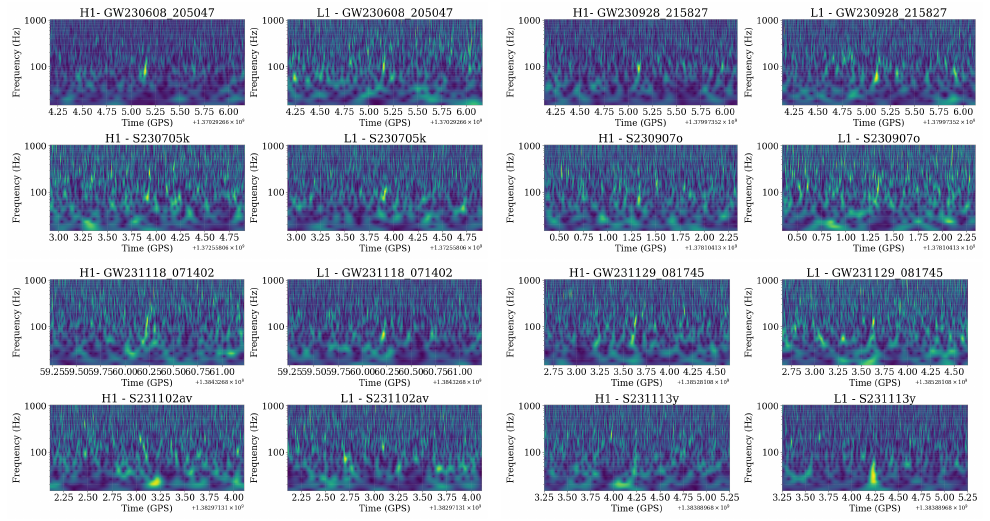}
    \caption{In this figure, we plot the spectrograms of the super-threshold and sub-threshold event pairs in the H1 and L1 detectors to visually inspect the frequency-evolution tracks of the signals underlying the data streams. These comparisons of spectrograms are shown for the event pairs GW230608\_205047 - S230705k (top left), GW230928\_215827 - S230907o (top right), GW231118\_071402 - S231102av (bottom left), and GW231129\_081745 - S231113y (bottom right) respectively.}
    \label{fig:spectrograms}
\end{figure*}

\begin{figure*}
    \centering
    \includegraphics[width=\linewidth]{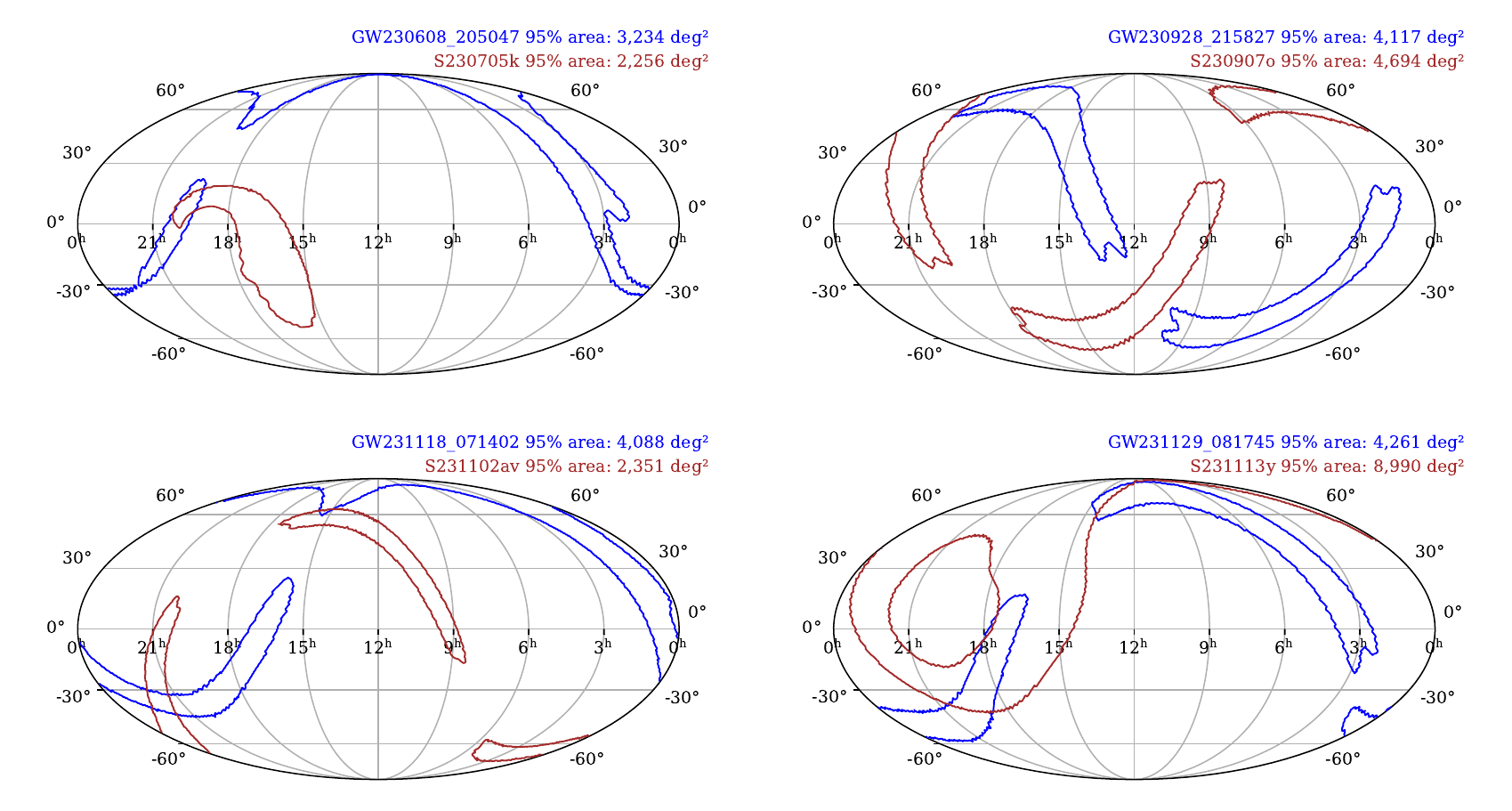}
    \caption{In this figure, we show the 95\% credible interval sky-maps of the interesting sky-map pairs, for which the cumulative cross-correlation SNR is $\geq 2$. We find that, GW230608\_205047 - S230705k do not overlap; GW230928\_215827 - S230907o and GW231118\_071402 - S231102av pairs have slight overlap in the sky and GW231129\_081745 - S231113y contain a significant fraction of overlaps. It is important to note that, these events being observed by H1-L1 detectors only, the sky-localization is quite unconstrained.}
    \label{fig:skymaps}
\end{figure*}

\begin{enumerate}
    \item We find the cumulative cross-correlation during a timescale of the timespan within which the signal spends its time in the detector (from 20Hz till the point of merger), for a cross-correlation timescale of $1/8$ s. These picks 4 events events with a threshold cross-correlation of 2, in the H1 and L1 detectors. We find four pairs of events for which the cumulative signal cross-correlation stands out from the noise cross-correlation above $2\sigma$. Chronologically, these event pairs are the following:
    \begin{enumerate}
        \item \href{https://gwosc.org/eventapi/html/GWTC-4.0/GW230608_205047/v1/}{GW230608\_205047} - \href{https://gracedb.ligo.org/superevents/S230705k/view/}{S230705k}.
        \item \href{https://gwosc.org/eventapi/html/GWTC-4.0/GW230928_215827/v1/}{GW230928\_215827} - \href{https://gracedb.ligo.org/superevents/S230907o/view/}{S230907o}.
        \item \href{https://gwosc.org/eventapi/html/GWTC-4.0/GW231118_071402/v1/}{GW231118\_071402} - \href{https://gracedb.ligo.org/superevents/S231102av/view/}{S231102av}.
        \item \href{https://gwosc.org/eventapi/html/GWTC-4.0/GW231129_081745/v1/}{GW231129\_081745} - \href{https://gracedb.ligo.org/superevents/S231113y/view/}{S231113y}.
    \end{enumerate}
    
    The cross-correlation between the event pair (top-panel) and the cross-correlation SNR (bottom-panel) is presented in the figure \ref{fig:snr_0.125s_16s}. With the color-bar, we show the astrophysical false alarm rate (FAR) of the sub-threshold event. The FAR of these events ranges from $\rm 1/day > FAR > 1/yr$.

    \item We inspect the spectrograms of these event pairs in the H1 and L1 detectors. These spectrograms of the super-events and sub-events are shown in the figure \ref{fig:spectrograms}, with a 2s window, with the starting and ending time of the window at $\pm 1$s  from the GPS time of the alert. Although this check is very qualitative in nature and has nothing to do with the detection technique, it is an irreducible part of the analysis to reject any correlated glitches or non-Gaussian noise artifacts from the data. These comparisons of spectrograms are shown for the event pairs GW230608\_205047 - S230705k (top left), GW230928\_215827 - S230907o (top right), GW231118\_071402 - S231102av (bottom left) and GW231129\_081745 - S231113y (bottom right), respectively. We observe that for S230705k, there is a coexisting feature in the spectrogram for both detectors, any frequency evolving feature is not well noticeable \footnote{This does not necessarily mean that no signal is present and only noise is there, as it might indicate a very short signal observed for the merger of very massive black holes.}. For S230907o, there is a frequency evolution in L1 but not in H1 \footnote{This does not necessarily mean that the signal is not astrophysical in nature, as the sky-position of the event may lie at the blind spot of the second detector.}. For the event S231102av, we cannot even distinguish any coexisting feature in either of the two detectors.  However, a sudden dump of energy is noticed in the H1 detector; however, the burst energy is not coherent among the detectors. For the event S231123y, we observe a coexisting feature in both H1 and L1, the signal is brighter in L1, but we cannot observe any frequency evolution track there. In contrast, the frequency evolution track is observable in the H1 detector, although it is dimmer there than in L1. Altogether, we do not observe any prominent event features among all these four sub-events; we would like to clarify that none of these events indicates any strong presence of correlated noise artifacts. 
    
    \item We calculated the 95\% credible interval (C.I.) regions of the sky-maps of these event pairs to check whether they have any overlap between their estimated sky-positions. The sky-maps were obtained from \href{https://gracedb.ligo.org/.}{GraceDb}. We find that GW230608\_205047 - S230705k do not contain any overlap; The pairs GW230928\_215827 - S230907o and GW231118\_071402 - S231102av have a nearly negligible overlap and the pair GW231129\_081745 - S231113y has a significant overlap. The skymaps of the event pairs and their overlaps are shown in the figure \ref{fig:skymaps}. We note that all events have 95\% CI sky-regions at an area $\rm \geq 1,000 deg^2$ since these events are observed with H1 and L1 detectors only, resulting in low constraining capability of the possible sky-position of the events in the sky.
    \item We calculate the duration of the signal in the band to find whether the GW cycles spend sufficient time within the observation period. We calculate the duration of the super-threshold signals to be within 0.124s, 0.116s, 0.154s and 0.148s for the super-events GW230608\_205047, GW230928\_215827, GW231118\_071402 and GW231129\_081745, respectively. Due to inaccessible frequency ranges and the short durations of the super-events spent within the detectors' frequency bands, the characterization of the signals to be lensed cannot be made with our current generation of detectors. 
    \item  As none of the pairs survive the previous tests, we do not need to infer the source properties of the sub-event. However, to have some insight about the GW source, if the signal exists at all, we perform a Bayesian parameter estimation through \texttt{BILBY} \citep{Ashton:2018jfp}, details of which can be found in the appendix \ref{app:pe_event}. 
\end{enumerate}

We find that there exists no event pair that meets the complete detection criterion. Although we obtained four pairs of events from the cross-correlation test alone, none of those events passed a statistically significant detection at $3\sigma$ in both H1 and L1 detectors. However, performing a further inspection on these pairs of events rejects their lensing hypothesis on the basis that the duration of the signals is not sufficient to make a claim. Therefore, from the whole super+sub-threshold GW catalog of a total of 885 events producing 11,207 pairs, we do not find even one single candidate for multi-image lensing. Throughout its strong rejection criteria for pairs of events, the framework displays the strength of the detection method as first proposed in the work \texttt{GLANCE} \citep{Chakraborty:2024net}.

\section{False lensing alarm rate for lensed events}\label{sec:far}

\begin{figure*}
    \centering
    \includegraphics[width=0.7\linewidth]{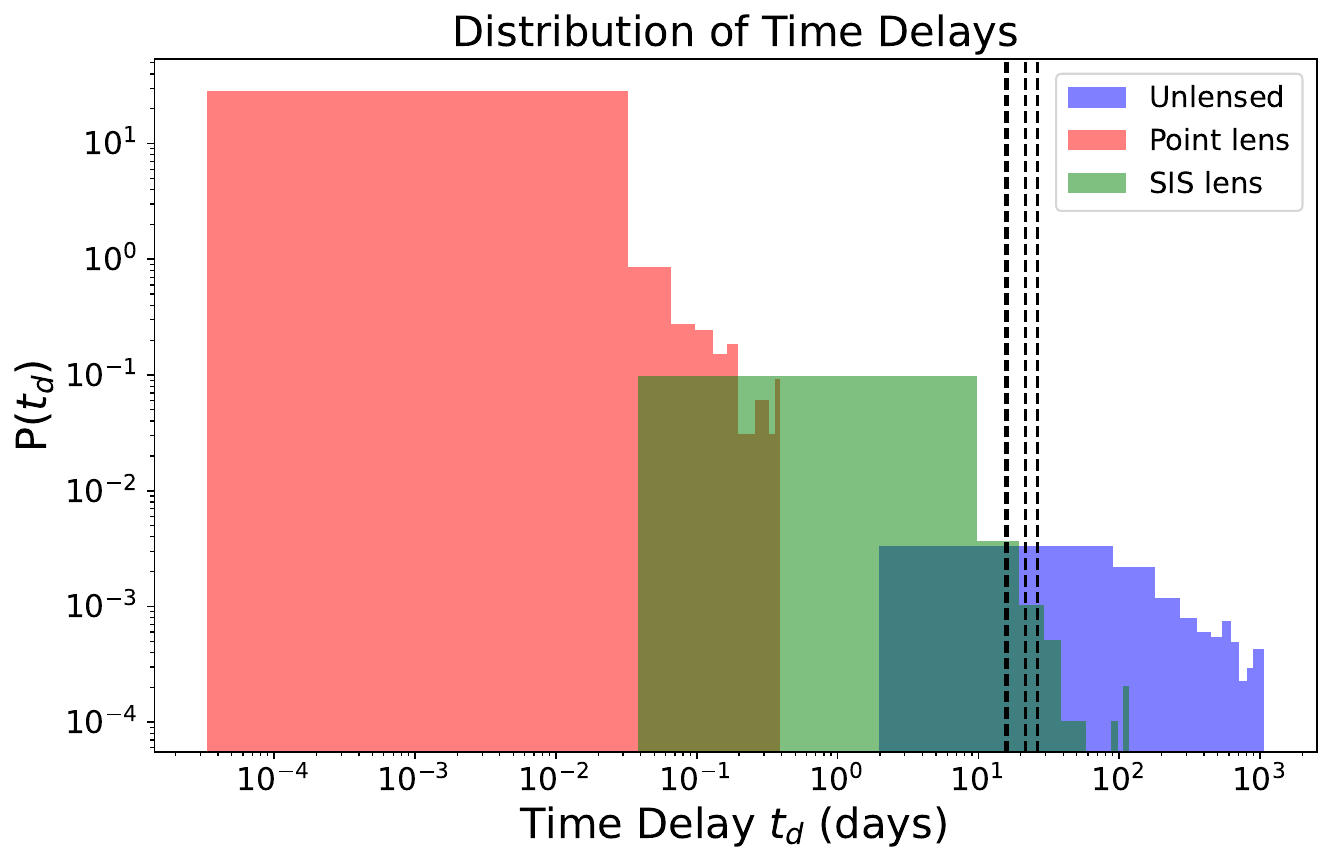}
    \caption{In this figure we show the time-delay distribution of lensed events coming from point mass lenses and SIS lenses. We also show the time delays of astrophysical unlensed GW events, with sky-overlapping pairs. We also show the time delays between the event pairs we picked up by cross-correlation searches at $\geq 2 \sigma$ with the vertical dashed lines. From left to right, these pairs are: GW231129\_081745 - S231113y, GW231118\_071402 - S231102av, GW230928\_215827 - S230907o and GW230608\_205047 - S230705k respectively.}
    \label{fig:td_dist}
\end{figure*}

To understand how the delay-times between candidate lensing pairs compare with the lensing time-delay from astrophysical objects and the false lensing mimickers (overlapping sky-position GW signals) \footnote{See the work \citep{chakraborty2026falsealarmratesdetecting} for a detailed analysis of the astrophysical false alarms for lensing}, we plot the distribution of allowed time-delays for various astrophysical systems and false lensing mimickers in the figure \ref{fig:td_dist}. In the same figure, vertical lines represent the time delay between the top four highest cross-correlation events. The masses of the point-mass lensing systems (representation of lensing by supermassive black hole systems) are simulated from a Schechter mass function \citep{1974ApJ...187..425P} with $\alpha= -1.5$ and $M_{*} = 10^{11} M_{\odot}$ in the mass range $[10^6, 10^9] M_{\odot}$. The models and the parameters chosen here are motivated by the following works \citep{Weigel_2016, McLeod_2021} 
The singular isothermal sphere lensing (representation of a massive galaxy embedded in a dark matter halo) masses are also coming from the Schechter mass function with the same parameters but in the range $[10^{9}, 10^{13}] M_{\odot}$. The dimensionless source impact parameter for both cases is obtained from a physically motivated distribution $p(y) \propto y$ up to a $y_{max}=1$. The time-delay distributions  obtained for these systems are presented in figure \ref{fig:td_dist}. Together with these lensing time-delay distributions, we obtain the time-delay distribution between sky-overlapping events appearing as false lensing alarms (as observed by two LIGO detectors in the US and the Virgo detector in Italy with the O4 noise sensitivities for an observation period of three years \citep{KAGRA:2013rdx, LIGOScientific:2014pky, PhysRevD.111.062002, LIGO:2024kkz, PhysRevD.102.062003, PhysRevLett.123.231107, PhysRevD.93.112004, PhysRevLett.116.131103, Harry:2010zz, VIRGO:2014yos, PhysRevLett.123.231108, Virgo:2022ysc, Luck:2010rt, 2014CQGra..31v4002A, Dooley:2015fpa, KAGRA:2020tym, PhysRevD.88.043007, Somiya:2011np})\footnote{The incorporation of Virgo contributes to the sky-overlap estimation in a significant way. Therefore, with Virgo offline in O4a, the sky positions estimated with H1 and L1 alone would be broader. However, the minimum time delay of astrophysical contaminants is of the order of the inverse of the detection rate of GW events. Therefore, the non-inclusion of Virgo would not drastically change the left edge of the unlensed lensing contaminant time-delay distribution shown in the figure \ref{fig:td_dist}. }. The events are generated from the observationally motivated mass and spin distribution from the LVK population analysis work from GWTC-4 \citep{theligoscientificcollaboration2025gwtc40populationpropertiesmerging}. This has been followed up from one of our previous works \citep{chakraborty2026falsealarmratesdetecting}, please refer to it for a more detailed description of the simulation. 

With the time-delay distributions, we present the time delay between the GW events found in this analysis in the figure \ref{fig:td_dist} with dashed vertical lines. These lines correspond to the time-delays between the event pairs GW231129\_081745 - S231113y, GW231118\_071402 - S231102av, GW230928\_215827 - S230907o and GW230608\_205047 - S230705k, respectively (from left to right). The time delay between these pairs does not correspond to supermassive black hole lensing; however, they can fit in the lensing scenario by more massive SIS lens systems. These events are also falling in the scenario where astrophysical coincidences can get confused with multiple images of the same event due to lensing.

\section{Constraints on the lensing rate}\label{sec:rate}
In this analysis, we do not observe any potential lensing candidate among super-threshold and sub-threshold events forming pairs. In our previous searches on wave-optics lensing from GWTC-3 \citep{Chakraborty:2025maj} and GWTC-4 \citep{Chakraborty:2025pxt} events with the residual cross-correlation technique \texttt{$\mu$-GLANCE}\citep{Chakraborty:2024mbr}, we found two candidates of interest. The event \href{https://gwosc.org/eventapi/html/GWTC-2.1-confident/GW190408_181802/v2/}{GW190408\_181802} from GWTC-3 showed a statistical significance of $ \geq 3\sigma$, although further tests do not show that the event is a strong candidate for lensing. In GWTC-4, we tested the lensing hypothesis on the most massive BBH merger GW event \href{https://gwosc.org/eventapi/html/GWTC-4.0/GW231123_135430/v2/}{GW231123\_135430} (often known as GW231123 only) which showed a statistical significance of $1.4\sigma -7.1\sigma$ depending on the GW waveform model, indicating a high degree of systematics associated with the event, and we remain inconclusive of the lensing status of the event. The findings from the lensing searches performed by LVK collaboration can be found here \citep{LIGOScientific:2023bwz, lvk_gwtc4_lensing}. However, if we assume these two events were unlensed, from the non-observation of lensed events from the total of 865 events until GWTC-4, then a forecast on the the maximum rate of detectable lensed events. The findings in our work lead to lensing at a chance lower than $\leq 1/865 \approx 0.00115 \%$ \footnote{It is important to note that these events are not the "detected" ones, in their literal sense, as these events did not meet the specific detection criteria conditioned on them.}. Given that these observations were made within a span of eight months (O4a), we can estimate the rate of such lensed events. We calculated that with the current sensitivity of the LVK detector network, we would expect $\leq 1.5$ lensed events/yr. The number agrees with the rate calculation in our previous work \citep{Chakraborty:2025pxt} and is in the same ballpark as previous lensing rate estimates \citep{Chen:2003uu, Robertson:2020mfh, Mukherjee:2021qam, Xu:2021bfn}.

\section{Summary and Conclusion}\label{sec:conclusion}

In this work, we performed the first model-independent lensing search with the super- and sub-threshold events jointly. We deployed the cross-correlation-based search technique \texttt{GLANCE} for the detection, as it is able to capture signals of the same morphology while suppressing uncorrelated noise artifacts. For a robust detection, we picked up the two detector events with the astrophysical false alarm rate $\geq 1 / yr$ as sub-threshold events along with the catalog events as super-threshold ones. We formed pairs between these events, forming around $\approx 11,000$ event pairs in total. We tentatively found four pairs of events lying $2\sigma$ away from the noise in both the H1 and L1 detector strains. Following up on these pairs with cumulative cross-correlation over time, spectrogram checks, and Bayesian parameter estimation, we find that none of the four pairs stand out as a potential lensing candidate. We calculate that the non-observation of lensed GWs would immediately put a constraint on the future detection of lensing, which we found to be $\leq$ 1.5/yr.

It is to be noted that the catalog of sub-threshold GW events started during O3; however, the number of sub-threshold events from O3 published in GraceDb is negligible \footnote{The number of sub-threshold events in the O3 run of LVK has merely $\approx 10$ events, see \url{https://gracedb.ligo.org}}. Therefore, this analysis relied on the O4a catalog events + sub-threshold ones. In O4a, only ground-based GW detectors H1 and L1 were observing \citep{2025arXiv250818079T}, allowing us to combine two detector combinations in this lensing search. In O4b, Virgo was operational; therefore, its observations would allow us to have one more independent channel of strain information with better sky-localization of sources. Therefore, future observation runs with an improved sensitivity Virgo detector \citep{KAGRA:2013rdx}, LIGO-India \citep{LIGO_India} and KAGRA \citep{KAGRA:2020tym, Somiya:2011np} would allow stronger claims of a potential lensing detection. In this regard, the next-generation ground-based and space-based GW detectors in the same band of observable frequencies, such as Cosmic-Explorer \citep{Punturo:2010zza, Reitze:2019iox}, Einstein Telescope \citep{Hild:2010id, LIGOScientific:2016wof}, and frequencies like deciHz e.g. TianGO \citep{Kuns:2019upi} or mHz e.g. LISA \citep{2021arXiv210801167B} would allow for many more GW detections in the coming days. With the current lensing observational rates of a few in a thousand, a detection is likely to happen in the coming years.  

\section{Acknowledgement}
The authors are thankful to Mick Wright for carefully reviewing the manuscript as part of the LIGO Publication and Presentation Policy and for providing valuable suggestions on the draft of the paper. This work is part of the ⟨Data$|$Theory⟩ Universe Lab, which is supported by the Department of Atomic Energy, Government of India. We acknowledge the support of the Department of Atomic Energy, Government of India, under Project Identification No. RTI 4012. This research is supported by the Prime Minister Early Career Research Award, Anusandhan National Research Foundation, Government of India. The authors express gratitude to the computer cluster of \texttt{⟨data|theory⟩ Universe-Lab} for computing resources used in this analysis. We thank the LIGO-Virgo-KAGRA Collaboration for providing noise curves \citep{H1_O1, L1_O1, H1_O2, L1_O2, H1_O3a, L1_O3a, H1_O3b, L1_O3b, O4_noise, advLIGO}. LIGO, funded by the U.S. National Science Foundation (NSF), and Virgo, supported by the French CNRS, Italian INFN, and Dutch Nikhef, along with contributions from Polish and Hungarian institutes. The research leverages data and software from the Gravitational Wave Open Science Center, a service provided by LIGO Laboratory, the LIGO Scientific Collaboration, Virgo Collaboration, and KAGRA. Advanced LIGO's construction and operation receive support from STFC of the UK, Max-Planck Society (MPS), and the State of Niedersachsen/Germany, with additional backing from the Australian Research Council. Virgo, affiliated with the European Gravitational Observatory (EGO), secures funding through contributions from various European institutions. Meanwhile, KAGRA's construction and operation are funded by MEXT, JSPS, NRF, MSIT, AS, and MoST. This material is based upon work supported by NSF’s LIGO Laboratory which is a major facility fully funded by the National Science Foundation. We acknowledge the use of the following python packages in this work: NUMPY \citep{harris2020array}, SCIPY \citep{2020SciPy-NMeth}, MATPLOTLIB \citep{Hunter:2007}, PYCBC \citep{alex_nitz_2024_10473621}, GWPY \citep{gwpy}, LALSUITE \citep{lalsuite}, BILBY \citep{Ashton_2019}, PESUMMARY \citep{Hoy:2020vys}, CORNER \citep{corner}.

\bibliography{biblio}

\appendix

\section{Usage of different timescales for cross-correlation analysis}\label{app:cc_timescale}

\begin{figure*}
    \centering
    \includegraphics[width=\linewidth]{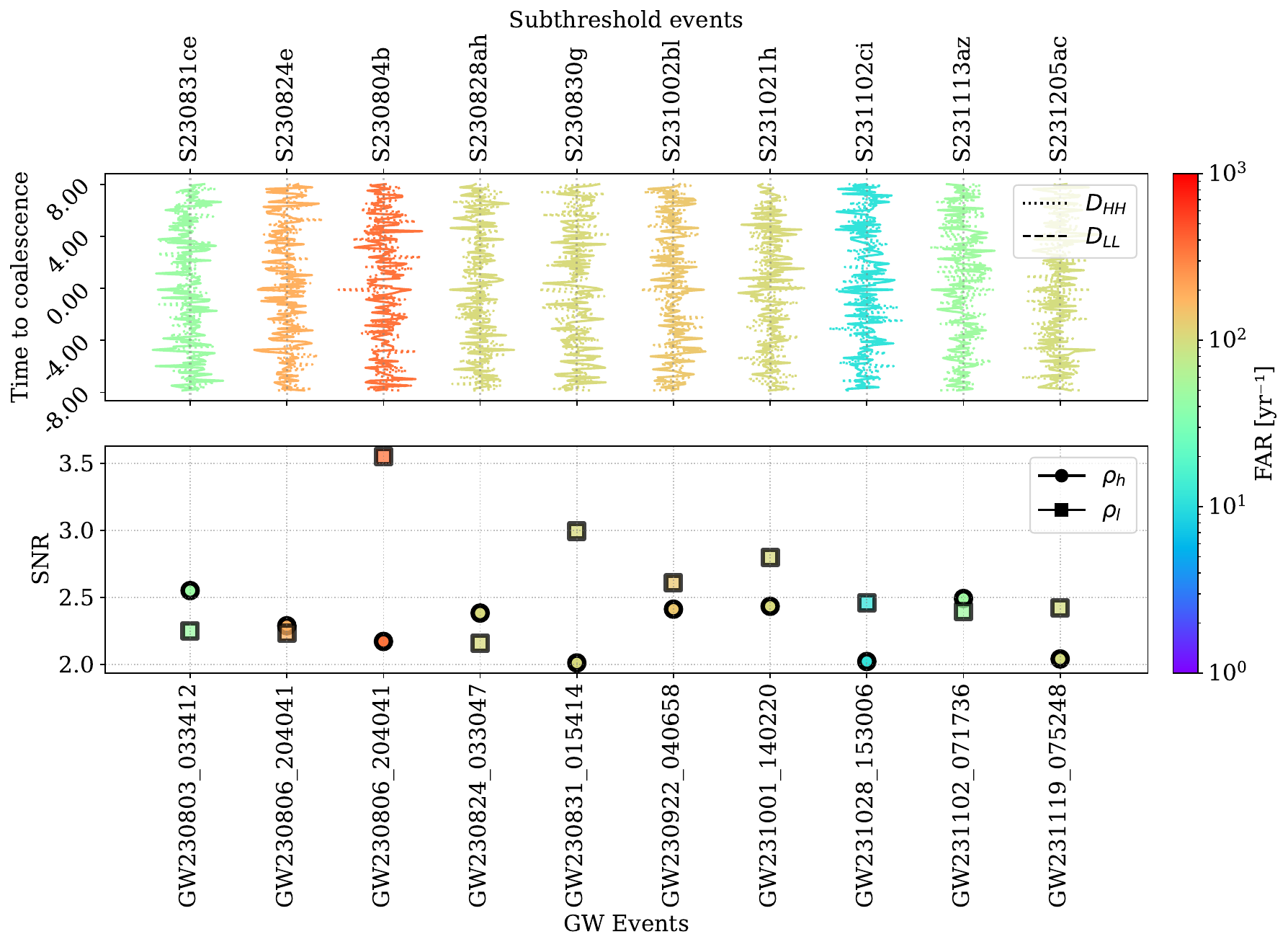}
    \caption{In this figure, we show the most important candidates in the cross-correlation lensing search for the cross-correlation timescale of 1/16s. We find 10 pairs above the two detector SNR of 2, the signals are short ($\leq 0.1s$), thus we reject their lensing hypothesis. }
    \label{fig:snr_0.0625s_16s}
\end{figure*}

We also used a cross-correlation timescale of 1/16 s to evaluate the correlation between the super-events and sub-events. We find that 10 events appear in the cross-correlation SNR $\geq 2$. However, all of these signals are again short in their time span spent within the LVK observation band. To be precise, all of these signals spend $\leq 0.1$ s within the observable band. Therefore, we reject the lensing hypothesis for all these events. However, for the reader's interest, we show the diagnostic plot for 1/16 s timescale in the figure \ref{fig:snr_0.0625s_16s}. The interesting pairs are shown in the figure, with super-events in the bottom panel and sub-events in the top panel. The false alarm rates are represented with the color bar on the right side.

\section{Bayesian inference of the sub-event sources}\label{app:pe_event}

To characterize the GW sub-events and  understand their source properties, we applied a Bayesian parameter estimation. We tried to estimate the following source parameters: the component masses (in the redshifted detector frame) ($\rm m_1$ and $\rm m_2$), the component spin magnitudes ($\rm a_1$ and $\rm a_2$) and their angles with respect to the orbital angular momentum ($\rm \theta_1$ and $\rm \theta_2$), and the following extrinsic parameters: luminosity distance to source ($\rm d_l$), inclination angle between the total angular momentum and the line of sight ($\rm \theta_{jn}$, time of coalescence ($\rm t_c$), polarization angle ($\rm \psi$), right ascension ($\rm ra$ and $\rm dec$), and the coalescence angle ($\rm \phi_c$). We find that the parameter estimations do not converge for any of the four events, whereas two of the runs get finished in $\approx$ 12 hr, the other two do not finish, given their 2 day+ of runtime. All of them show poor constraints on the masses, with bi-modality and uniform posteriors for the events. To mention the specification of the runs, the PE was performed over 16s of data chunks that encompassed the events. The minimum and maximum frequencies  are chosen to be 20 Hz and 512 Hz respectively. The chosen waveform model is \texttt{IMRPhenomXPHM}\citep{Pratten:2020ceb} with priors on the parameters as follows. The prior on chirp mass $\mathcal M_c$ is chosen to be uniform within [5, 200] M$_\odot$, the mass-ratio $q$ prior is [0.1, 1], the luminosity distance $\rm d_L$ prior is [10, 20000]Mpc, and the time of coalescence ($\rm t_c$) prior is kept at $[\rm t_{event}-0.25s, t_{event}+0.25s]$\footnote{Here, the $\rm t_{event}$ prior is the alert time as provided by GraceDb.}, all other parameters' priors are chosen as full ranges with uniform priors. The live points were kept at 1000 and the stopping criteria were kept at $\rm d(logZ)\leq 1$ ($\rm Z$ be the evidence of hypothesis for a GW signal in the data with \texttt{IMRPhenomXPHM} model), which is a comparatively looser condition of sampling convergence.

\end{document}